\documentclass[12pt,thmsa]{article}
\usepackage{amssymb}
\usepackage{graphicx}

\input{tcilatex}
\begin{document}

\begin{center}
{\large Generation of maximally entangled states of two cavity
modes}

~

 {F. Saif$^{1,2,}$\footnote[4]{E-mail:
saif@fulbrightweb.org}, M. Abdel-Aty$^{3,}$, M. Javed$^2$ R.
Ul-Islam$^2$ and A.-S. F. Obada}

~

{\small $^1$Department of Physics, University of Arizona, Tucson
85721, Arizona, USA
\newline $^2$Department of Electronics,
Quaid-i-Azam University, 45320 Islamabad, Pakistan
\newline
$^3$Mathematics Department, Faculty of science, Sohag University,
82524 Sohag, Egypt; Mathematics Department, College of Science,
Bahrain University, 32038 Kingdom of Bahrain }

~

{\bf  Appl. Math. Inf. Sci. Vol. 1 No. 3 pp. 323 (2007)}

\end{center}

In this letter we present a scheme for generating maximally
entangled states of two cavity modes which enables us to generate
complete set of Bell basis states having rather simple initial state
preparation. Furthermore, we study the interaction of a two-level
atom with two modes of electromagnetic field in a high Q cavity. The
two-level atom acts as a control qubit and the two mode
electromagnetic field serves as a target qubit. This simple system
of quantum electrodynamics provides us experimentally feasible
universal quantum logic gates.

~


~

 Quantum entanglement is a striking nonclassical property of
quantum composite systems. In addition to the conceptual problems of
reality and locality in quantum physics, quantum entanglement talks
about technological aspects of quantum communications, quantum
computation and quantum cryptography [1]. \smallskip There exist
situations in which the transition between the upper and lower
levels of an atom is mediated by two photons when the energy
separation between the levels is close to double the photon
frequency. This process and its multiphoton counterparts are
important because they can be used to study statistical properties
of the optical field [2]. It is therefore desirable to investigate
the nonlinear interaction of the three-level atom with the cavity
field in the case of the two-photon resonance transitions~[21].

\textrm{Our goal in} this letter is to discuss the possibility of
obtaining a scheme for generating maximally entangled states in this
model and to develop quantum universal logic gates which may
generate bimodal entanglements. We have shown that it is possible to
generate Bell-type states having rather simple initial state
preparation. To reach our goal we have to find the exact time
dependent expressions for the final state of the system. This can be
achieved by solving the Schr\"{o}dinger equations of motion. In
order to realize our suggested scheme in laboratory experiment
within microwave region, we may consider slow $Rb$ atoms in higher
Rydberg states which have life time of the order of few milliseconds
[3]. These slow atoms, initially pumped to high Rydberg state, pass
through a high-Q superconducting cavity of dimension of a few
centimeters with a velocity of around 400 $m/s$ [3,4]. The
interaction time of the atom with different cavities can be
controlled by using a velocity selector and applying Stark field
adjustment in different cavities in order to make the atom resonant
with the field for a right period of time [3].

In quantum computation the basic key component is a universal logic
gate, which leads to any operation on a qubit. The quantum universal
logic gate
comprises quantum CNOT gate together with single-qubit Hadamard gate ~\cite%
{seven,sevena,eighta,eighta1}. In addition a quantum phase gate
serves as a universal quantum logic gate as well~\cite{eightb}. The
quantum controlled-NOT gate has been realized experimentally in
Ramsey atomic interferometry, by means of the selective driving of
optical resonances of two qubits undergoing a dipole-dipole
interaction ~\cite{ten}, and by the Bragg scattering of
atoms~\cite{eleven}. In this paper we suggest that the interaction
of a circular Rydberg atom with a high Q superconducting cavity
leads to develop quantum universal logic gate in the electromagnetic
field modes. The atom is in resonance with the two modes in the
presence and in the absence of Stark field. Thus the Rydberg atom
acts as a control qubit whereas the two mode elecrtomagnetic field
provides target qubit.

Generation of maximally entangled states between two parties implies
existence of the two parties in two different states with equal
probability~[5]. The generation of maximally entangled states
between two cavity modes implies conditional existence of photons in
either of the two modes~[21]. The maximal entangled state may appear
as one photon in one cavity mode and the other mode in vacuum, or as
the two cavity modes are either in vacuum state or have one photon
each with equal probability. Such maximally entangled states
constitute EPR Bell states which provide a complete set of bases. In
order to develop the EPR Bell bases we prepare a three level atom in
$V$-configuration, such that, the two upper levels are in
superposition state. Hence, in order to generate two electromagnetic
field modes entanglement the interaction time of the atom with the
cavity is odd integer multiple of half of the Rabi cycles, $\lambda
_{1}$ and $\lambda _{2}$. This ensures that the cavity obtains one
photon in either of the two modes when atom is detected in ground
state after its propagation through the cavity field modes. This
implies that we find an entangled state between the two cavity
modes. Preparation of the atom in its upper state and letting it
interact for an interaction time equivalent to one fourth of Rabi
cycle leads to generate the other two Bell bases. As the atom
interacts with the first mode for a quarter of Rabi cycle it finds a
laser pulse in resonant with the other transition. Later an
interaction for a time half of the Rabi cycle provides another
entangled state of the cavity modes where either the two modes are
in vacuum or each having one photon. The two kind of entangled
states together make complete Bell bases.

We propagate a three-level atom through a cavity which contains
initially
field modes in vacuum. We may express the three levels as $|a\rangle $, $%
|b\rangle $, and $|c\rangle $ with their eigen energies as $E_{a}$,
$E_{b}$, and $E_{c}$. The dipole transition between the upper two
levels, $|a\rangle $ and $|b\rangle $, of the atom is forbidden,
whereas transitions from the two upper levels to lower level,
$|c\rangle $ are allowed. We consider that frequencies, $\omega
_{A}$ and $\omega _{B}$, of the two modes, $A$ and $B$,
respectively, of cavity field are in resonance with the transition
frequencies, such that $\omega _{A}=(E_{a}-E_{c})/\hbar $ and
$\omega _{B}=(E_{b}-E_{c})/\hbar $. With the help of a Ramsey field
we prepare the upper two levels of the atom in linear superposition
before it enters the cavity field. We may express the initial state
of the system as,
\begin{equation}
\left\vert \psi ^{(0)}\right\rangle =\left[ \cos \theta |a\rangle
+\sin \theta e^{i\phi }\Vert b\rangle \right] |0_{A},0_{B}>,
\end{equation}%
where, $\phi $ is the relative phase between two atomic states.

We write the interaction picture Hamiltonian in the dipole and
rotating wave approximation as
\begin{equation}
H=\hbar g_{1}\left( a\left| a\right\rangle \left\langle c\right|
+a^{\dagger }\left| c\right\rangle \left\langle a\right| \right)
+\hbar g_{2}\left( b\left| b\right\rangle \left\langle c\right|
+b^{\dagger }\left| c\right\rangle \left\langle b\right| \right) ,
\end{equation}
where $g_{1}$ and $g_{2}$ are vacuum Rabi frequencies of the two
modes while $a(a^{\dagger })$ and $b(b^{\dagger })$ are the
annihilation(creation) operators of the two cavity modes $A$ and
$B$, respectively. The atom-field state vector can be written as
\begin{eqnarray}
\left| \psi ^{(t)}(A,B)\right\rangle &=&\cos \theta \cos
g_{1}t\left| a,0,0\right\rangle -i\cos \theta \sin g_{1}t\left|
c,1,0\right\rangle
\nonumber \\
&+&e^{i\phi }\sin \theta \cos g_{2}t\left| b,0,0\right\rangle
-ie^{i\phi }\sin \theta \sin g_{2}t\left| c,0,1\right\rangle .
\end{eqnarray}
For the generation of maximally entangled field state between two
cavity modes such that if one mode has one photon then the other
will be in vacuum, the atom after its interaction with the cavity
fields, is required to be detected in ground state $\left|
c\right\rangle $. This leads to the condition that probability
amplitudes of the states $\left| c,1,0\right\rangle $ and $\left|
c,0,1\right\rangle $ are equal, i.e., $\sin g_{1}t=\sin g_{2}t$, for
$\theta =\pi /4$ and $\phi =0$. The total
probability of detecting the atom in ground state $P_{c}$ is determined as $%
P_{c}=\left( \sin ^{2}g_{1}t+\sin ^{2}g_{2}t\right) /2$. This
probability
becomes maximum when the time of interaction of atom with mode $A$ and mode $%
B$ is $m\pi /2g_{1}$ and $n\pi /2g_{2}$, respectively. Here, $m$ and
$n$ are odd integer numbers. Hence, in order to generate two mode
entanglement the time of interaction of the atom with the cavity is
odd integer multiple of half of the Rabi cycle. This ensures that
the cavity will obtain one photon in either of the two modes when
atom is detected in ground state after its propagation through the
cavity.

The interaction times of the atom with the two modes of the cavity
field would be different because of the different coupling constants
of each mode of radiation field. These interaction times of atom in
the cavity can be controlled by using a velocity selector before the
cavity and then applying Stark field adjustment so that atom becomes
resonant with the cavity field modes only for the suggested period
of time in each mode of the cavity field
[3]. Hence the atom passing in the superposition of levels $|a\rangle $ and $%
|b\rangle $ interacts with the two cavity modes $A$ and $B$ for
$m\pi /2g_{1} $ and $n\pi /2g_{2}$ interaction times, respectively.
As a result the atom leaves the cavity in ground state and develops
entangled states between the two cavity modes, viz.,
\begin{equation}
\left| \psi (A,B)\right\rangle =\frac{-i}{\sqrt{2}}\left[ \left|
0_{A},1_{B}\right\rangle \pm e^{i\phi }\left| 1_{A},0_{B}\right\rangle %
\right] ,
\end{equation}
by controlling the interaction time of the atoms with the cavities.

In order to generate the other two Bell bases we prepare the three
level atom, in level $\left\vert a\right\rangle $ without
considering the
superposition of upper two levels, $\left\vert a\right\rangle $ and $%
\left\vert b\right\rangle $. We let the excited atom pass through
two cavities successively, which are prepared initially in vacuum
state. The transition from level $\left\vert a\right\rangle $ to
$\left\vert c\right\rangle $ is again in resonance with cavity mode
$A$, whereas the transition from $\left\vert b\right\rangle $ to
$\left\vert c\right\rangle $ is in resonance with cavity mode $B$.
We adjust the interaction time of the atom with first cavity field
such that it sees a $\pi /2$ pulse. Hence,
there occurs equal probability of finding the atom in ground state $%
\left\vert c\right\rangle ,$ after contributing one photon in the
cavity mode $A,$ and of finding the atom in the excited state
$\left\vert a\right\rangle ,$ leaving the cavity mode in vacuum
state. With a proper choice for the atomic dipole phase [6], we find
an atom-field entanglement, such that,
\begin{equation}
\left\vert \psi _{at}(A,B)\right\rangle =\frac{1}{\sqrt{2}}\left[
\left\vert a,0_{A}\right\rangle +\left\vert c,1_{A}\right\rangle
\right] \otimes \left\vert 0_{B}\right\rangle .
\end{equation}%
Before the atom enters the next cavity, we apply a laser field
resonant to atomic transition $\left\vert b\right\rangle $ to
$\left\vert c\right\rangle $. The width of the beam is adjusted such
that the exiting atom from the first cavity field in the ground
state $\left\vert c\right\rangle ,$ is pumped to excited state
$\left\vert b\right\rangle $ with unit probability. However, if the
exiting atom is in excited state $\left\vert a\right\rangle $ after
interacting with the cavity mode $A$, the laser field will provide
no excitation to the atom.

After passing through the laser field, the atom interacts with cavity mode $%
B $, which is initially in vacuum state. The interaction time of the
atom
with the field is adjusted such that the atom in the excited state $%
\left\vert b\right\rangle $ will be detected in ground state
$\left\vert
c\right\rangle $ with unit probability adding a photon in the cavity mode $B$%
. However, if the atom enters the cavity in the excited state
$\left\vert a\right\rangle $, it will contribute no photon and will
exit in the same atomic state, leaving the cavity mode $B$ in the
vacuum state. Let the atom exiting the second cavity interact with
another field, resonant with the atomic transition $|a\rangle
\rightarrow $ $|c\rangle $. The interaction time of $\pi /2$ and
relative phase $\phi =\pi /2$ lead to $|a\rangle
\rightarrow \left(\cos\theta |a\rangle -\sin \theta|c\rangle \right) $ and $%
|c\rangle \rightarrow \left( \cos\theta|a\rangle +\sin
\theta|c\rangle \right)$ [7]. This makes the final state as
\begin{eqnarray}
\left\vert \psi _{af}(A,B)\right\rangle &=&\left( \cos \theta
\left\vert 0_{A},0_{B}\right\rangle +\sin \theta \left\vert
1_{B},1_{B}\right\rangle
\right) \otimes |a\rangle  \nonumber \\
&&+\left( \sin \theta \left\vert 1_{B},1_{B}\right\rangle -\cos
\theta \left\vert 0_{A},0_{B}\right\rangle \right) \otimes |c\rangle
.
\end{eqnarray}
Detection of the atom in state $|a\rangle $ or $|c\rangle $ and choosing $%
\theta =\pi /4,$ generates the entangled states of cavity field modes $\frac{%
1}{\sqrt{2}}\left( \left\vert 0_{A},0_{B}\right\rangle +\left\vert
1_{B},1_{B}\right\rangle \right) $ or $\frac{1}{\sqrt{2}}\left(
\left\vert 1_{B},1_{B}\right\rangle -\left\vert
0_{A},0_{B}\right\rangle \right) .$ Therefore, we find the
entanglement of the two modes of radiation field with the atomic
states, for a time larger than the life time of the propagating
atom, as
\begin{equation}
\left\vert \psi (A,B)\right\rangle =\frac{1}{\sqrt{2}}\left[
\left\vert 1_{B},1_{B}\right\rangle \pm \left\vert
0_{A},0_{B}\right\rangle \right] , \label{ent}
\end{equation}%
which describes the entanglement of the two cavity field modes. The
interaction times of the atoms with the cavity field mode $A$, laser
field,
and cavity field mode $B$ are found as $m\pi /4g_{1},$ $\pi /\Omega ,$ and $%
n\pi /2g_{2},$ respectively, where $m$ and $n$ are odd integers. \textrm{%
Here }$\Omega $\textrm{\ is the Rabi frequency of the laser field
which interacts with the atom between the two cavities.} If the
relative difference of interaction times of atoms with the two
cavities is $\pi $, we
generate the entangled state with negative sign, given in equation~(\ref{ent}%
). Hence, we can obtain the complete set of Bell basis by
controlling the interaction times of the atom with the cavities in
both schemes.

It should also be pointed out that our analysis assumes a perfectly
isolated system. In a real experiment, the quantum motion of a
trapped atom is obviously limited by sources of decoherence. In
order to realize the suggested scheme in a laboratory experiment
within the microwave region, we may consider slow Rb atoms in higher
Rydberg states, which have lifetimes of the order of a few
milliseconds. Finally, one may say that, the atomic decay rates,
interaction times, and cavity lifetime ensure that the atom does not
decay spontaneously. As this entanglement remains only for the
cavity lifetime period, any application regarding this entangled
state should be accomplished during this period [5].

We conclude our analysis by saying that in the present letter we
have presented a powerful and useful tool to the generation of
maximally entangled states using a three-level atom interacting with
a bimodal cavity. We have demonstrated that, in many standard
concepts and experiments in quantum optics where it appears
necessary to use bimodal cavity, it is equally as valid to generate
complete set of Bell basis states using sources of vacuum states.

In order to develop quantum universal logic gates able to generate
maximum entanglement in the two cavity field modes, we consider a
two-level circular Rydberg atom, which passes through a high Q
superconducting cavity. The cavity contains nondegenerate
orthogonally polarized modes $M_A$ and $M_B$ with mode frequencies
$\omega_{A}$ and $\omega_{B}$. We consider the atomic transition
frequency, $\omega_{eg}$ $(= \omega_{e}- \omega_{g})$, in
resonance with the field frequency $\omega_{A}$. Here, $\omega_{e}$ and $%
\omega_{g}$ are the frequencies associated with excited state
$|e\rangle$ and with the ground state $|g\rangle$, respectively. In
presence of an
electric field the excited state $|e\rangle$ changes to the state $%
|e^{\prime}\rangle$ and the atomic transition frequency
$\omega_{e^{\prime}g} $ becomes equal to the electromagnetic field
frequency $\omega_{B}$ due to the Stark effect~\cite{A}. Thus the
atom emits a photon coherently in the cavity mode $M_{B}$ at a
different frequency, $\omega_{B}$, in the presence of Stark field.
Latter, the final state of the atom is analyzed in a state-selective
field ionization detector. We show that by controlling the coherent
interaction it is possible to realize two qubit quantum CNOT logic
gate and a single Hadamard gate in the system.

The quantum Controlled NOT logic gate is a two-input two-output
logic gate
which requires a control qubit $\left|q_1\right\rangle $ and target qubit $%
\left|q_2\right\rangle $. The state of the control qubit $\mid
q_1\rangle $ controls the state of the target qubit, $\mid
q_2\rangle $, such that
\begin{equation}
\mid q_1\rangle \mid q_2\rangle \ \rightarrow \mid q_1\rangle \mid
q_1\oplus q_2\rangle ,  \nonumber
\end{equation}
where, $\oplus $ indicates addition modulo 2 ~\cite{eighta1}. This
implies that the the target qubit is flipped if the control qubit
carries logic one and remains unchanged if the control qubit carries
logic zero.

As discussed above, in our study we take the control qubit as a two
level atom, which is defined in a two dimensional Hilbert space with
$|e\rangle$ and $|g\rangle$ as basis vectors. Here, $\left|
e\right\rangle $ expresses the excited state of the two level atom,
and $\left| g\right\rangle$ indicates its ground state.

The two non-degerate and orthogonally polarized cavity modes $M_{A}$ and $%
M_{B}$ make the target qubit, $\left|q_2\right\rangle $. The target
qubit is
defined in two dimensional Hilbert space spanned by the state vector $%
|\psi_1\rangle=|1_A, 0_B\rangle$, which expresses the presence of
one photon
in mode $A$, when no photon is present in mode $B$, and the state vector $%
|\psi_2\rangle=|0_A, 1_B\rangle$, which indicates that the mode $A$
is in vacuum state, when one photon is present in the mode $B$.

Interaction of the two level atom, acting as a control qubit, with
the electromagnetic cavity containing two field modes, acting as a
target qubit, leads to the universal two-qubit control NOT logic
gate and single bit Hadamard gate. We prepare the two level atom in
the ground state, $\left| g\right\rangle $, in a Ramsey cavity. It
enters an electromagnetic cavity
which contains single photon of the field mode, $M_{A}$, whereas the mode $%
M_{B}$ is in vacuum state. The transition frequency of the atom is
taken equal to the frequency of the mode $M_{A}$, hence, the atom
interacts resonantly with the field.

The interaction of the two level atom with the electromagnetic field mode, $%
M_A$ is described by the standard Jaynes-Cummings interaction Hamiltonian~%
\cite{A}, expressed as
\begin{equation}
V=\hbar \mu_1(a^{\dagger }\sigma + \sigma^{\dagger}a),  \nonumber
\end{equation}
where $a^{\dagger}$ $(a )$ is field creation (annihilation) operator, $%
\sigma^{\dagger}=\left|e\right\rangle \left\langle g\right |$ ($%
\sigma=\left| g\right\rangle \left\langle e\right |$) is atomic
raising (lowering) operator, and $\mu_1$ is the coupling constant.
Hence the atom-field combined state of the system becomes,
\begin{equation}
|\psi\rangle=c_g|g, 1_A\rangle+ c_e|e, 0_A\rangle.
\end{equation}
The probability amplitudes, $c_g$ and $c_e$, which govern the
evolution of the atom initially in its ground state, change as a
function of interaction time $t$ such that,
\begin{equation}
c_{g}\left( t\right) =\cos \left(\Omega _{A}t/2\right) ,
\end{equation}
and
\begin{equation}
c_{e}\left( t\right) =-i \sin \left(\Omega _{A}t/2\right) .
\end{equation}
The frequency, $\Omega_{A}=2\mu_1\sqrt{n_A}$, describes the Rabi
frequency of the atom in the mode $M_{A}$ containing $n_A$ number of
photons. The atom interacts for a time $\pi/\Omega_{A}$ with mode
$M_A$ and completes half of the Rabi oscillation. As a result it
absorbs the cavity photon in mode $M_{A} $ and jumps to its excited
state, $\left| e\right\rangle $.

After the interaction time $\pi/\Omega_{A}$, we apply a Stark field
which
shifts the excited state from $\left| e\right\rangle$ to $\left| \acute{e}%
\right\rangle$. Hence the atom observes a change in its transition
energy and finds itself resonant with the mode, $M_{B}$. The atom
interacts with the cavity mode, $M_{B}$, resonantly and follows the
interaction Hamiltonian, viz.,
\begin{equation}
V=\hbar \mu_2(b^{\dagger }\hat\sigma+ \hat\sigma^{\dagger}b)
\nonumber
\end{equation}
where $b^{\dagger }$ $(b)$ is creation (annihilation) operator of
the field mode $M_B$, $\hat\sigma^{\dagger}=\left|e\right\rangle
\left\langle g\right|$ $(\sigma=\left| g\right\rangle \left\langle
e\right|)$ atomic raising (lowering) operator, and $\mu_2$ is the
coupling constant. The probability amplitudes of the resonant atom
flips as,
\begin{equation}
c_{e^{}}\left( t\right) =\cos \left(\Omega _{B}(t-t_0)/2\right) ,
\end{equation}
and
\begin{equation}
c_{g}\left( t\right) = -i \sin \left(\Omega _{B}(t-t_0)/2\right),
\end{equation}
where $t>t_0=\pi/\Omega_{A}$ and $\Omega_B=2\mu_2\sqrt{n_B+1}$,
where $n_B$ describes the number of photons in cavity $B$. After an
interaction with the mode $M_B$ for a time
$\pi(\Omega_A+\Omega_B)/\Omega _{A}\Omega_B$, the atom leaves the
cavity in its ground state $\left| g\right\rangle $ and thus
contributes one photon to the cavity mode $M_{B}$. Therefore, the
atom performs a swapping of electromagnetic fields between two field
modes by a controlled interaction.

In case we prepare the two level atom in its ground state, $\mid
g\rangle $, when the electromagnetic cavity contains the mode
$M_{A}$ in its vacuum state and a single photon in electromagnetic
field mode, $M_{B}$. The atom becomes resonant with the
electromagnetic cavity field mode, $M_{B}$, in the presence of the
Stark field at time, $t=0$. It exhibits a controlled interaction
with the field mode $M_B$, for a time $\pi /\Omega_{B}$ which is
equal to half of the Rabi oscillation time. Here, $\Omega_B=2\mu_2\sqrt{n_B}$%
. Thus the state of the system at $t=\pi /\Omega_{B}$ becomes $%
|e^{},0,0\rangle$. We switch off the Stark field and let the atom
interact resonantly with mode $M_{A}$ for a time $\pi(\Omega_A+\Omega_B)/%
\Omega_A\Omega_B$, where $\Omega_A=2\mu_1\sqrt{n_A+1}$. Therefore,
the atom again leaves the cavity in its ground state
$\left|g\right\rangle $, and
performs field swapping by contributing one photon to the field mode, $M_{A}$%
.

The target qubit made up of the electromagnetic fields remains
unchanged if the control qubit, that is the two level atom is
initially in its excited state. The atom prepared in its excited
state $\left| e\right\rangle $ interacts with the mode $M_{A}$
containing one photon field. The Rabi oscillation frequency is given
by $\Omega_{A} = 2\mu_1\sqrt{n_A+1}$ thus it completes one Rabi
oscillation in time $2\pi/\Omega_{A}=\sqrt{2}\pi/\mu_1$. The atom
leaves the cavity in state, $\left| e\right\rangle $, without
disturbing the field mode, $M_{A}$. However, if the cavity field
mode $M_{B}$ is in Fock state one, the atom becomes resonant with
$M_{B}$ in the presence of a Stark field to introduce zero detuning
between atomic transition frequency and the field frequency. Hence,
the atom completes one Rabi oscillation in the field for the
interaction time $\sqrt{2}\pi/\mu_2$ and leaves the cavity in its
excited state $| e\rangle$ without contributing to the radiation
field mode, $M_{B}$.

Hadamadard gate generates superposition state of a qubit provided
the system is in one of the bases vectors of the two dimensional
Hilbert space. Hence, we consider the control qubit in state $\left|
g\right\rangle $ and target qubit in the electromagnetic field state
$|\psi_1\rangle=\left| 1_{A}, 0_{B}\right\rangle$. The atom
interacts with the cavity for a time equal to one fourth of Rabi
oscillation period, that is $\pi/2\Omega_A$. Thus the combined
atom-field state of the system becomes, $(\left| e, 0_{A},
0_{B}\right\rangle +\left| g, 1_{A}, 0_{B}\right\rangle )/\sqrt{2}$.
Later, we introduce a Stark field, which shifts the excited state
from $\left| e\right\rangle $ to $\left| e^{}\right\rangle $. Thus
the atom interacts in resonance with the mode $M_{B}$. The atom
leaves the cavity after time $\pi/\Omega_B$, in its ground state,
$\left| g\right\rangle $, and contributes one photon to the field in
mode, $M_{B}$. The resultant state of the system becomes a
superposition state of $\left| \psi_1\right\rangle $ and $\left|
\psi_2\right\rangle $, with an equal probability to find the system
in either of the two bases vectors.

In this paper we present a scheme to engineer two-qubit quantum
controlled NOT logic gate and single bit Hadamard gate by a
controlled interaction between two-mode, high Q, electromagnetic
cavity field and a two-level atom. For the purpose we take the
two-level atom as control qubit, whereas the target qubit is made up
of two modes of the cavity field. We express the development of
quantum controlled NOT logic gate in Table $1$, as,

\begin{eqnarray*}
\begin{tabular}{||l||l||l||l||}
\hline\hline $\left| q_1\right\rangle $ & $\left| q_2\right\rangle $
& $\left| q_1\right\rangle $ & $\left| q_1\oplus q_2\right\rangle $
\\ \hline\hline $\left| g\right\rangle $ & $\left| 1_{A},
0_{B}\right\rangle $ & $\left| g\right\rangle $ & $\left| 0_{A},
1_{B}\right\rangle $ \\ \hline\hline $\left| g\right\rangle $ &
$\left| 0_{A}, 1_{B}\right\rangle $ & $\left| g\right\rangle $ &
$\left| 1_{A}, 0_{B}\right\rangle $ \\ \hline\hline $\left|
e\right\rangle $ & $\left| 1_{A}, 0_{B}\right\rangle $ & $\left|
e\right\rangle $ & $\left| 1_{A}, 0_{B}\right\rangle $ \\
\hline\hline $\left| e\right\rangle $ & $\left|0_{A},
1_{B}\right\rangle $ & $\left| e\right\rangle $ & $\left|0_{A},
1_{B}\right\rangle $ \\ \hline\hline
\end{tabular}%
\end{eqnarray*}

In order to realize our scheme in laboratory, we may fellow the
experimental setup of the Ref.~\cite{twelve}. We consider circular
Rydberg rubidium atom, with principal quantum number $51$ and $50$
acting as levels $\left| e\right\rangle $ and $\left| g\right\rangle
$, respectively, as control bit. The transition from level
$|e\rangle $ to $| g\rangle $ occurs at frequency 51.1 GHz. A very
small rate of injection makes the probability of having two atoms at
the same time in the cavity very small. The optical cavity is a
Fabry-Perot resonator made of two spherical nibium mirrors. The two
orthogonally polarized TEM$_{900}$ modes, $M_{A}$ and $M_{B}$ have
the same Gaussaian geometry with waist 6 mm. The frequency splitting
occurs due to a slight mirror shape anisotropy. The photon damping
times are $T_{r,a}=1$ $ms$
and $T_{r,b}=0.9$ $ms$ for the electromagnetic field modes $M_{A}$ and $%
M_{B},$ respectively.

The engineering of the universal logic gate helps us to generate
multi-qubit quantum gates~\cite{ekert} and to engineering
entanglement, in various modes of cavity field~\cite{twelve,21}.
Furthermore, the experimental accessability of these logic gates
make it possible to implement quantum algorithms in the QED setup.
We may use the system for secret data communication i.e,
cryptography, and to perform teleportation.

\[
\]


\end{document}